\documentclass[%
 reprint,
unsortedaddress,
 amsmath,amssymb,
 aps,
]{revtex4-1}

\usepackage{graphicx}
\usepackage{dcolumn}
\usepackage{bm}

\usepackage{siunitx}
\usepackage[version=4]{mhchem}
\usepackage{amsmath}
\usepackage[colorlinks,bookmarksopen,bookmarksnumbered,citecolor=blue,urlcolor=blue,anchorcolor=blue,linkcolor=blue]{hyperref}

\begin{document}


\title{Tunable magnetic interactions in \ce{LaAlO3}/\ce{SrTiO3} heterostructures by ionic liquid gating}

\author{Chunhai Yin, Kumar Prateek, Wouter Gelling and Jan Aarts}
\affiliation{
	Huygens-Kamerlingh Onnes Laboratory, Leiden University, P.O. Box 9504, 2300 RA Leiden, The Netherlands\\
}

\date{\today}

\begin{abstract}
The gating effect achieved by an ionic liquid and its electric double layer allows for charge transfer which can be an order of magnitude larger than with conventional dielectrics. However, the large charged ions also causes inevitable Coulomb scattering in the conducting channel formed at the interface, which can limit the carrier mobility enhancement. In this work, we study the effect of the \ce{LaAlO3} thickness on the transport properties in \ce{LaAlO3}/\ce{SrTiO3} heterostructures by ionic liquid gating. We find that the transport properties of the \ce{LaAlO3}/\ce{SrTiO3} interface are dominated by the intrinsic interactions rather than the \ce{LaAlO3} thickness and possible effects from the ions in the liquid. We observe a Kondo effect, which is enhanced while increasing the gate voltage. We also observe a gate-tunable and temperature-dependent anomalous Hall effect, which always emerges near the Kondo temperature. Our experiments pave the way to manipulate the various magnetic interactions in \ce{LaAlO3}/\ce{SrTiO3} heterostructures.
\end{abstract}

\maketitle
The two-dimensional electron system (2DES) at the interface between \ce{LaAlO3} (LAO) and \ce{SrTiO3} (STO) \cite{ohtomo2004} has been the focus of intensive studies due to its intriguing properties, such as superconductivity \cite{reyren2007}, signatures of magnetism \cite{brinkman2007nm,ariando2011nc,lee2013NM} and even their coexistence \cite{li2011,bert2011}. It has also been demonstrated that the 2DES properties can be modulated by electrostatic gating. Pioneering works have been performed in the back-gate configuration, where the dielectric insulator is the STO substrate. Novel gate-tunable effects, such as insulator to metal transition \cite{thiel2006}, insulator to superconductor transition \cite{caviglia2008N} and Rashba spin-orbit coupling \cite{shalom2010PRL,caviglia2010PRL} have been reported. However, in this configuration high voltages of tens to hundreds of volts are required to achieve a sizable gating effect. Moreover, the high voltages also induce unavoidable modifications to the defect landscape in the STO substrate \cite{2019trap}. The top-gate geometry works in a similar way as the back-gate counterpart, but the insulating dielectric is the LAO overlayer. Compared with back-gating, the big advantage of top-gating is that it requires very low voltages to achieve sizeable gating effects. However, the device fabrication process is very complicated which is due to the requirement of multiple aligned lithography steps \cite{smink2017PRL}.

Electric double layer transistors (EDLTs) of the top-gate configuration provide an alternative approach.
In recent years, EDLTs have generated considerable research interest due to the capability of inducing high carrier densities in materials. EDLTs have been used for inducing superconductivity in STO \cite{ueno2008nm} and \ce{KTaO3} \cite{ueno2011nn} as well as for tuning transport properties in LAO/STO heterostructures \cite{lin2014ami,zeng2016acsnano,chen2016nl}. EDLTs use ionic liquids (ILs) as gate dielectrics, in which the size of the ions is usually larger than the lattice parameter of perovskite oxides. For instance, the average diameter of the ions in a widely used IL, called DEME-TFSI \cite{textil}, is $\sim$ \SI{0.7}{\nano\meter} \cite{kim2005jtes}. When the electrolyte touches the channel, the long-range Coulomb potentials created by the ions will cause inevitable scattering of the conduction electrons and therefore limit the mobility enhancement, as was shown for experiments on BN-encapsulated graphene \cite{petach2017acsnano}. In a similar fashion, \citet{gallagher2015nc} have recently reported a ten-fold improvement of the carrier mobility in STO single crystals by covering the channel with a thin BN flake.

In LAO/STO heterostructures, the LAO film is a natural protection for the channel. However, one would expect that there is a trade-off between obtaining high carrier densities, which requires thin LAO layers, and achieving high mobilities, which requires thick LAO layers. In this work, we study the effect of LAO film thickness on the transport properties of three LAO/STO EDLTs. We obtain the largest tunability in a device with 4 unit cells (uc) of LAO. We observe a gate-tunable Kondo effect in all the devices. More interestingly, we also observe a temperature-dependent nonlinearity of the Hall resistance, which is always accompanied by the emergence of Kondo effect. We attribute this anomaly to an anomalous Hall effect. Our results imply that the transport properties of LAO/STO EDLTs are mainly governed by the intrinsic interactions, while the LAO thickness has a minor effect.

\begin{figure}[t]
	\centering
	\includegraphics[width=\linewidth]{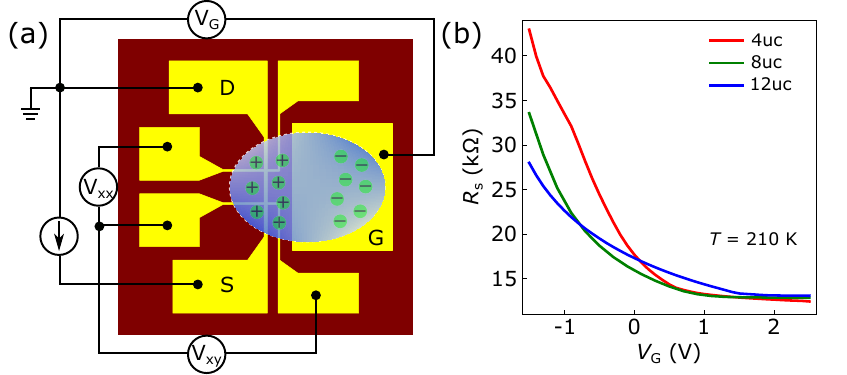}
	\caption{(a) A schematic of the EDLT device. Source and drain are labeled as S and D. The longitudinal ($V\rm_{xx}$) and transverse voltages ($V\rm_{xy}$) are measured simultaneously. The gate voltage ($V\rm_{G}$) is applied between the gate electrode (G) and the drain. (b) Sheet resistance ($R\rm_{s}$) as a function of $V\rm_{G}$ at \SI{210}{\kelvin} for the three devices with different thickness of the LAO layer as indicated. }
	\label{figRV}
\end{figure}

A schematic of the EDLT device is shown in Fig. \ref{figRV}(a). First, a sputtered amorphous AlO$_{x}$ hard mask in form of a negative Hall bar geometry (thickness $\sim$\SI{15}{\nano\meter}) was fabricated on a \ce{TiO2}-terminated STO (001) substrate by photolithography. Then, an LAO film was deposited at \SI{800}{\degreeCelsius} in an Ar pressure of \SI{0.04}{\milli\bar} by \SI{90}{\degree} off-axis sputtering \cite{yin2019prm}. Finally, the sample was \textit{in situ} annealed at \SI{600}{\degreeCelsius} in \SI{1}{\milli\bar} of oxygen for \SI{1}{\hour}. The length between two voltage probes is \SI{500}{\micro\meter}, and the Hall bar width is \SI{50}{\micro\meter}. It is well known that the critical thickness of LAO which gives rise to a conducting interface is 4 uc \cite{thiel2006}. We fabricated three devices with different thickness of LAO layer. Since we do not have thickness monitoring capability during growth, we used the calibrated growth rate of \SI{4.3}{\AA/\minute} \cite{yin2019prm} to estimate the thickness. Growth times were 3.5, 7.0 and 10.5 min, respectively, yielding the devices with thicknesses of 1.5, 3.0 and \SI{4.5}{\nano\meter}, which we refer to as 4uc, 8uc and 12uc \cite{textsm}. After sample fabrication, a droplet of IL, DEME-TFSI (IOLITEC), was applied onto the device surface covering the gate electrode and the channel.

\begin{figure*}[t]
	\centering
	\includegraphics[width=\linewidth]{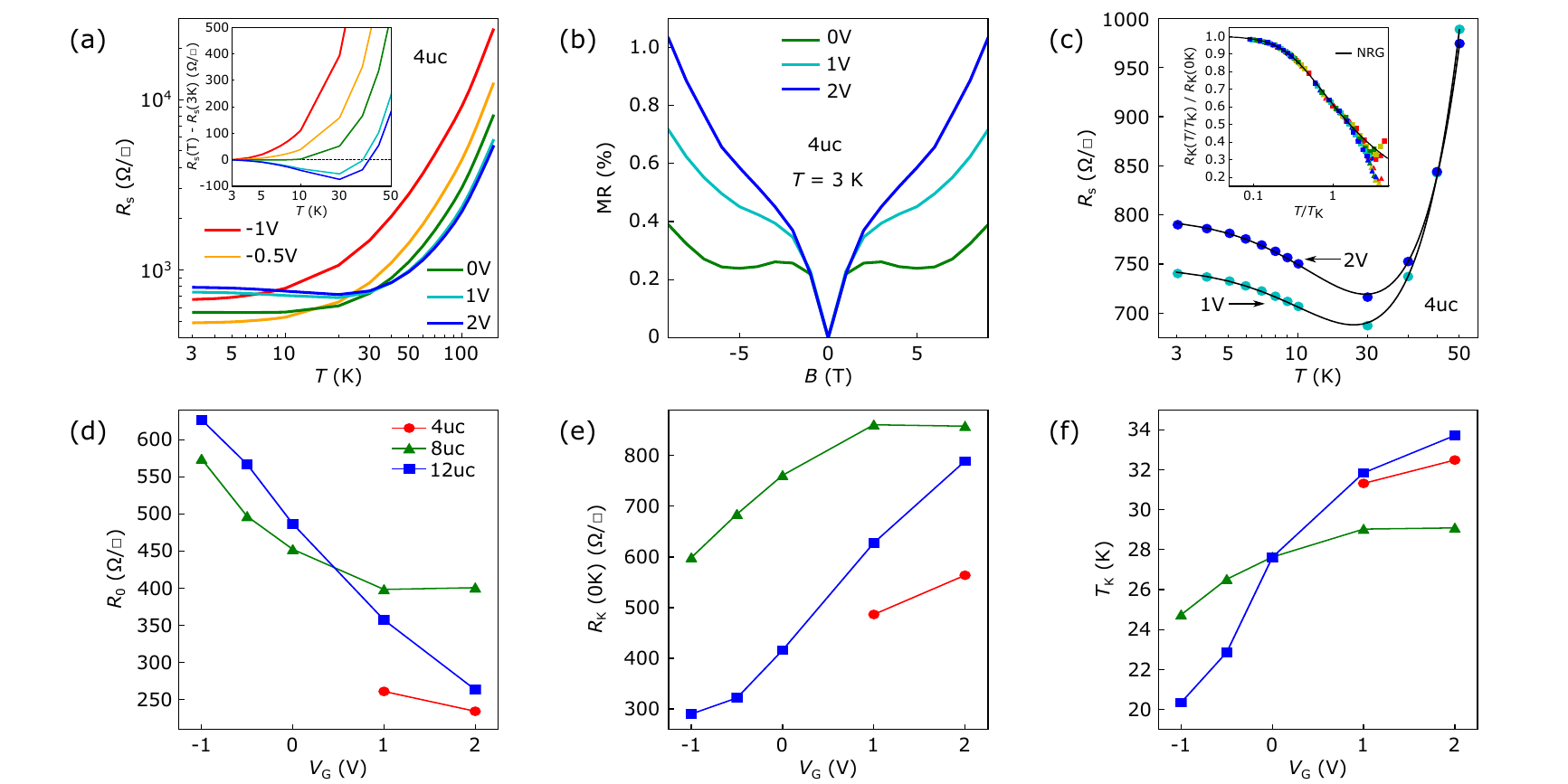}
	\caption{(a) $R\rm_{s}$ as a function of temperature at various $V\rm_{G}$s of device 4uc. The inset shows the resistance relative to its value at \SI{3}{\kelvin}, $R\rm_{s}$($T$)$-$$R\rm_{s}$(\SI{3}{\kelvin}). (b) Magnetoresistance (MR) at various $V\rm_{G}$s of device 4uc at \SI{3}{\kelvin}. (c) $R\rm_{s}$ minimum observed from 3 to \SI{50}{\kelvin} of device 4uc at \SI{1}{\volt} and \SI{2}{\volt}. The black lines are the Kondo effect fits using Eqs. (\ref{eqkondo}) and (\ref{eqRk}). The inset shows the normalized Kondo resistance ($R{\rm_{K}}(T/T{\rm_{K}})$/$R{\rm_{K}}(\SI{0}{\kelvin})$) versus the reduced temperature ($T/T{\rm_{K}}$) curves of the three devices collapse into a single black line, which is obtained from the numerical renormalization group (NRG) method. (d)-(f) Extracted Kondo parameters, (d) the residual resistance ($R_{0}$), (e) the Kondo resistance at zero temperature ($R{\rm_{K}}(\SI{0}{\kelvin})$) and (f) the Kondo temperature ($T{\rm_{K}}$), as a function of $V\rm_{G}$ of the three devices. }
	\label{figkondo}
\end{figure*}

The magnetotransport measurements were performed in a Physical Property Measurement System (Quantum Design) at temperatures down to \SI{3}{\kelvin} and magnetic fields up to \SI{9}{\tesla}. As shown in Fig. \ref{figRV}(a), a Keithley 6221 current source was used to apply a direct current ($I\rm_{DC}$ = \SI{1.0}{\micro\ampere}). Two Keithley 2182A nanovoltmeters were used to measure the longitudinal ($V\rm_{xx}$) and transverse voltages ($V\rm_{xy}$). The gate voltage ($V\rm_{G}$) was applied by a Keithley 2400 source meter. The freezing point of DEME-TFSI is $\sim$\SI{180}{\kelvin}. As a general procedure, we applied $V\rm_{G}$ at \SI{210}{\kelvin} and then waited for $\sim$\SI{15}{\minute} for the IL to equilibrate. The measurements were started at \SI{150}{\kelvin} and the leakage current was less than \SI{1.0}{\nano\ampere} during the measurements.

Fig. \ref{figRV}(b) shows the sheet resistance ($R\rm_{s}$) as a function of $V\rm_{G}$ at \SI{210}{\kelvin} for the three devices. As expected, the gate tuning effect on $R\rm_{s}$ decreases as the LAO thickness increases. Above $V\rm_{G}$ = \SI{1.5}{\volt}, $R\rm_{s}$ almost saturates in all of the devices. We find that the gate voltage sweeps are nicely reproducible, which allows us to exclude any possible electrochemical reaction occurring on LAO surface \cite{ueno2008nm}. The forward and backward sweeps are also reversible, indicating that no electron trapping \cite{2019trap} is induced in the top-gate configuration.

\begin{figure*}[t]
	\centering
	\includegraphics[width=\linewidth]{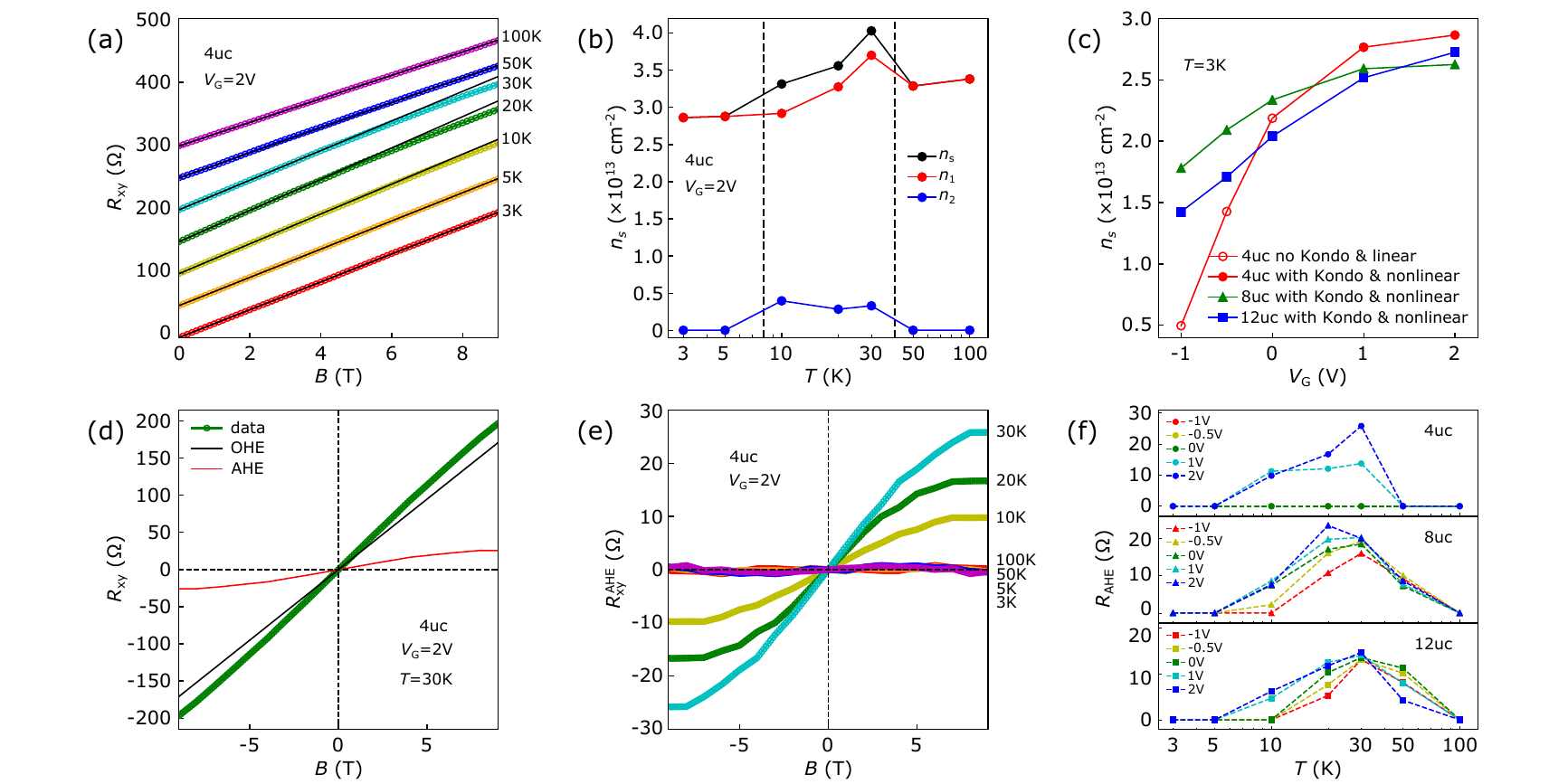}
	\caption{(a) Hall resistance ($R\rm_{xy}$) at various temperatures of device 4uc at $V\rm_{G}$ = \SI{2}{\volt}. The curves are separated by an offset of \SI{50}{\ohm} and the black lines are linear fits to $R\rm_{xy}$. (b) Temperature dependence of the carrier density ($n\rm_{s}$) of device 4uc at $V\rm_{G}$ = \SI{2}{\volt}. The values are extracted by fitting the magnetotransport data with a two-band model. $n\rm_{1}$ and $n\rm_{2}$ represent the carrier densities of different conduction channels and their sum is $n\rm_{s}$. (c) $n\rm_{s}$ as a function of $V\rm_{G}$ of the three devices at \SI{3}{\kelvin}. (d) An example for the determination of OHE and AHE terms from $R\rm_{xy}$ measured on device 4uc at \SI{30}{\kelvin} and $V\rm_{G}$ = \SI{2}{\volt}. (e) Deduced AHE as a function of magnetic field at various temperatures of device 4uc at $V\rm_{G}$ = \SI{2}{\volt}. (f) Anomalous Hall resistance ($R\rm_{AHE}$) as a function of temperature at various $V\rm_{G}$s of the three devices.}
	\label{fighall}
\end{figure*}

First, we present the temperature dependence of $R\rm_{s}$ measured under various values of $V\rm_{G}$. Fig. \ref{figkondo}(a) shows the data of device 4uc \cite{textsm}. At $V\rm_{G}$ = \SI{0}{\volt}, $R\rm_{s}$ first decreases from \SI{150}{\kelvin} to \SI{10}{\kelvin} and then remains unchanged. In the negative voltage regime, $R\rm_{s}$ decreases monotonously to \SI{3}{\kelvin}, but in the positive voltage regime, $R\rm_{s}$ first reaches a minimum at $\sim$ \SI{30}{\kelvin} and then increases as temperature decreases to \SI{3}{\kelvin}. For a better view, the resistance relative to its value at \SI{3}{\kelvin}, $R\rm_{s}$$(T)$$-$$R\rm_{s}$(\SI{3}{\kelvin}), is plotted in the inset of Fig. \ref{figkondo}(a). In earlier reports, this resistance upturn has been attributed to either a weak localization (WL) effect \cite{caviglia2008N,wong2010prb} or a Kondo effect \cite{lee2011prlKondo,lin2014ami}. Fig. \ref{figkondo}(b) shows the magnetoresistance (MR) of device 4uc measured at \SI{3}{\kelvin} and under various $V\rm_{G}$s \cite{textsm}. The MR curves exhibit sharp cusps in low magnetic fields, which is the characteristic feature of a weak antilocalization (WAL) effect \cite{seiler2018PRB} and allows us to exclude the WL effect.

In a metallic system where itinerant electrons couple antiferromagnetically with localized moments, spin-flip scattering gives rise to a resistance minimum, which is known as the Kondo effect \cite{kondo1964}.
The Kondo effect is characterized by a Kondo temperature, $T{\rm_{K}}$. For $T$ $\textless$ $T{\rm_{K}}$, the resistance increases logarithmically and eventually saturates at low temperatures. The $R\rm_{s}$$(T)$ can be described by a simple formula \cite{lee2011prlKondo}
\begin{equation}
R(T) = R_{0} + aT^{2} + bT^{5} + R{\rm_{K}}(T/T{\rm_{K}}),
\label{eqkondo}
\end{equation}
where $R_{0}$ is the residual resistance, and the $T^{2}$ and $T^{5}$ terms represent the temperature-dependent electron-electron and electron-phonon interactions, respectively. $R{\rm_{K}}(T/T{\rm_{K}})$ is the universal resistivity function with the empirical form
\begin{equation}
R{\rm_{K}}(T/T{\rm_{K}}) = R{\rm_{K}}(\SI{0}{\kelvin})\Big(\frac{1}{1+(2^{1/s}-1)(T/T{\rm_{K}})^{2}}\Big)^{s},
\label{eqRk}
\end{equation}
where $R{\rm_{K}}(\SI{0}{\kelvin})$ is the Kondo resistance at zero temperature and $s=0.225$. Fig. \ref{figkondo}(c) shows the fitting results of device 4uc at $V\rm_{G}$ = \SI{1}{\volt} and \SI{2}{\volt}. The extracted values of $R_{0}$, $R{\rm_{K}}(\SI{0}{\kelvin})$ and $T{\rm_{K}}$ as a function of $V\rm_{G}$ for the three devices are plotted in Fig. \ref{figkondo}(d)-(f).

Numerical renormalization group (NRG) method normalizes the Kondo contribution to the transport by a universal parameter $T/T{\rm_{K}}$ \cite{bulla2008rmp}. It is often used as a criterion to verify the Kondo scenario. As shown in the inset of Fig. \ref{figkondo}(c), $R{\rm_{K}}(T/T{\rm_{K}})$/$R{\rm_{K}}(\SI{0}{\kelvin})$ as a function of $T/T{\rm_{K}}$ measured under various $V\rm_{G}$s of the three devices all collapse into a single curve, which confirms that the Kondo effect is the origin of the observed resistance minimum \cite{textsm}. As can be seen from Fig. \ref{figkondo}(e)-(f) in particular, $R{\rm_{K}}(\SI{0}{\kelvin})$ and $T{\rm_{K}}$ are a function of the gate voltage. Such a gate-tunable Kondo effect has been previously observed in undoped STO single crystals \cite{lee2011prlKondo}, LAO/STO \cite{lin2014ami} and $\gamma$-\ce{Al2O3}/STO \cite{niu2017nl} EDLTs, and is consistent with our results. It is noteworthy to mention that the Kondo effect is caused by the interaction between itinerant electrons and localized electrons \cite{lee2011prlKondo} rather than with conventional magnetic impurities, such as iron. In the presence of magnetic impurities, one would expect to observe a stronger Kondo effect at a lower carrier density, not the opposite trend seen here.

Next, we present the temperature-dependent nonlinearity of the Hall resistance ($R\rm_{xy}$) measured under various $V\rm_{G}$s. Fig. \ref{fighall}(a) shows $R\rm_{xy}$ as a function of magnetic field ($B$) of device 4uc measured at various temperatures and at $V\rm_{G}$ = \SI{2}{\volt} \cite{textsm}. The $R\rm_{xy}$($B$) curves are separated by an offset of \SI{50}{\ohm} for clarity and the black lines are their linear fits. It can be seen that the $R\rm_{xy}$($B$) curves are linear at high temperatures (above \SI{50}{\kelvin}) and low temperatures (below \SI{10}{\kelvin}), but nonlinear at intermediate temperatures. We find that the emergence of nonlinear $R\rm_{xy}$($B$) is always accompanied by the observation of Kondo effect. For the cases of device 4uc at $V\rm_{G}$ = \SI{-1}{\volt}, \SI{-0.5}{\volt} and \SI{0}{\volt}, $R\rm_{xy}$($B$) curves are linear at all temperatures.

One possibility for a nonlinear Hall effect is two-band transport \cite{joshua2012NC,smink2017PRL}. The band structure of the LAO/STO interface is formed by the \ce{Ti} $t_{2g}$ orbitals. For LAO films grown on STO (001) substrates, the $d_{xy}$ band stays below the $d_{xz,yz}$ bands in energy \cite{salluzzo2009PRL,santander2011nature,smink2017PRL}. In the one-band regime, only the $d_{xy}$ band is populated. The carrier density ($n\rm_{s}$) and mobility ($\mu$) can be determined by \(n_\textrm{s}=B/eR_\textrm{xy}\) and \(\mu=R_\textrm{xy}/BR_\textrm{s}\), where $e$ is the electron charge. Tuning the Fermi level to above the bottom of the $d_{xz,yz}$ bands triggers a Lifshitz transition, resulting in a nonlinear $R\rm_{xy}$($B$). In the two-band regime, the carrier densities and mobilities can be extracted by fitting the magnetotransport data with a two-band model \cite{smink2017PRL}
\begin{equation}
\sigma_{xx} = e\sum_{i=1,2}\frac{n_{i}\mu_{i}}{1+(\mu_{i}B)^{2}}; \quad \sigma_{xy} = eB\sum_{i=1,2}\frac{n_{i}\mu_{i}^{2}}{1+(\mu_{i}B)^{2}},
\end{equation}
where $n\rm_{i}$ and $\mu\rm_{i}$ are the carrier density and mobility of the $i$th band.

Fig. \ref{fighall}(b) plots the carrier density as a function of temperature of device 4uc at $V\rm_{G}$ = \SI{2}{\volt} as an example. It can be seen that the total carrier density ($n_\textrm{s}=n_\textrm{1}+n_\textrm{2}$) is \SI{2.86e13}{\centi\meter^{-2}} at \SI{3}{\kelvin} and increases to \SI{4.03e13}{\centi\meter^{-2}} at \SI{30}{\kelvin}, corresponding to an increase of $\sim$ \SI{41}{\percent}. In earlier reports, such a nonlinear Hall effect at intermediate temperatures has been only reported in a 26 uc LAO/STO sample \cite{guduru2013prb}, where the occupation of the $d_{xz,yz}$ bands has been attributed to thermally excited electrons. However, we argue that this cannot serve as a convincing explanation for our results. First, when the 2DES enters the nonlinear regime, we obtain a large increase of $n\rm_{s}$, ranging from \SI{22}{\percent} to \SI{41}{\percent}, in our three devices at various $V\rm_{G}$s. But in Ref. \cite{guduru2013prb}, the increase is only $\sim$ \SI{3}{\percent}, which is much less than our case. Second, for the cases of device 4uc at $V\rm_{G}$ = \SI{-1}{\volt}, \SI{-0.5}{\volt} and \SI{0}{\volt}, we find that $n\rm_{s}$ remains almost constant at all temperatures. If thermal excitation indeed gives rise to additional electrons, one would expect to observe an increase of $n\rm_{s}$ under all $V\rm_{G}$s. Third, in earlier reports, the critical carrier density ($n\rm_{L}$) window for Lifshitz transition is 1.5$-$\SI{2.9e13}{\centi\meter^{-2}} \cite{joshua2012NC,smink2017PRL,2019trap}. Within our measurement limit, we observe linear $R\rm_{xy}$($B$) at \SI{3}{\kelvin} in all of the devices. Fig. \ref{fighall}(c) shows the extracted $n\rm_{s}$(\SI{3}{\kelvin}) as a function of $V\rm_{G}$. The threshold carrier density for the emergence of Kondo effect differs among devices, but the overall trend is the same. Most of the values are in the $n\rm_{L}$ window, which indicates that a higher magnetic field is required to reliably analyze the transport regime. However, even the 2DES is in the two-band regime, we should expect to see the nonlinearity appearing in magnetic fields higher than \SI{9}{\tesla}, rather than the extraordinary behavior observed here.

\begin{figure}[t]
	\centering
	\includegraphics[width=\linewidth]{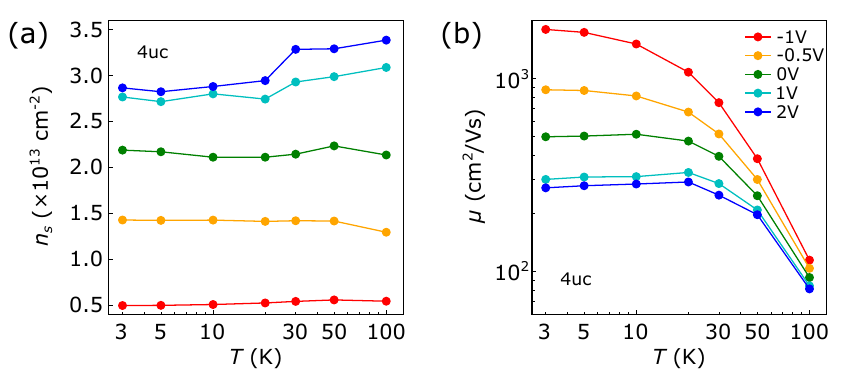}
	\caption{(a) Temperature dependence of (a) carrier density ($n_{s}$) and (b) mobility ($\mu$) of device 4uc at various $V\rm_{G}$s. The values are obtained by fitting $R{\rm_{xy}^{OHE}}$.}
	\label{fignmu}
\end{figure}

An alternative possibility for the nonlinear Hall effect is the anomalous Hall effect (AHE), which occurs in solids with ferromagnetic order \cite{nagaosa2010rmp}. In a system with AHE, $R\rm_{xy}$ is consisted of the ordinary Hall effect (OHE) term and the AHE term. As shown in Fig. \ref{fighall}(d), the OHE term, $R{\rm_{xy}^{OHE}}$, can be obtained from a linear fit to the high field limit of $R\rm_{xy}$. Subtracting $R{\rm_{xy}^{OHE}}$ from $R\rm_{xy}$ will give the AHE term, $R{\rm_{xy}^{AHE}}$ \cite{stornaiuolo2016nm}. Fig. \ref{fighall}(e) shows the $R{\rm_{xy}^{AHE}}$ terms obtained at various temperatures for device 4uc at $V\rm_{G}$ = \SI{2}{\volt}. Similar to common observations in the LAO/STO system, the $R{\rm_{xy}^{AHE}}$ curves show no hysteresis. \citet{gunkel2016prx} have proposed a phenomenological Langevin-type function to describe the $R{\rm_{xy}^{AHE}}$. However, we could not obtain a satisfactory fit due to the lack of a clear magnetization saturation in the limited magnetic field. \citet{christensen2018np} have recently reported a non-saturating AHE even up to \SI{15}{\tesla}. Fig. \ref{fighall}(f) shows the anomalous Hall resistance ($R{\rm_{AHE}}$) taken from the high field limit of $R{\rm_{xy}^{AHE}}$, which increases as $V\rm_{G}$ is increasing. Similar temperature-dependent AHE effect has been reported in STO single crystals \cite{lee2011prlAHE} and LAO/STO:Fe heterostructures \cite{zhang2018prb}. Fig. \ref{fignmu}(a) and \ref{fignmu}(b) plot the $n_{s}$ and $\mu$ as a function of temperature of device 4uc under various $V\rm_{G}$s. Device 8uc and 12uc show similar behavior with less tunability \cite{textsm}.

The Kondo effect originates from antiferromagnetic coupling between itinerant electrons and localized spins \cite{kondo1964}. In our LAO/STO system, the Kondo scattering centers are localized and unpaired electrons \cite{lee2011prlKondo}. The density of these localized electrons increases when increasing $V\rm_{G}$. The temperature range in which AHE emerges is closely related to $T\rm_{K}$. At low temperatures, $T$ $\ll$ $T\rm_{K}$, the localized spins are effectively screened by the continuum states of the metal host upon forming coherent many-body singlet states \cite{nozires1974jltp}. A rough estimation of the energy scale, $g$$\mu\rm_{B}$$B$ $\approx$ $k\rm_{B}$$T\rm_{K}$ ($g$ is the g factor, $\mu\rm_{B}$ is the Bohr magneton and $k\rm_{B}$ is the Boltzmann constant), indicates that a magnetic field of several tens of Tesla is required to break down the Kondo screening. Therefore a linear $R\rm_{xy}$ is observed. At high temperatures, $T$ $\gg$ $T\rm_{K}$, the alignment of localized moments by magnetic field is countered by thermal fluctuations, resulting in a linear $R\rm_{xy}$ as well. At intermediate temperatures, $T$ $\sim$ $T\rm_{K}$, the polarization of localized spins by magnetic field could give rise to a ferromagnetic ordering, resulting in the anomalous Hall effect. The origin of this non-hysteretic magnetization is still unclear. It may be due to spiral magnetic ordering \cite{banerjee2013np} or polarization of unpaired spins \cite{cheng2015n,pai2016arxiv}.

In conclusion, we have studied the effect of LAO film thickness on the transport properties of LAO/STO EDLTs, and obtained the largest tunability in terms of carrier density and mobility in the device with 4 uc of LAO. We have shown a gate-tunable Kondo effect which originates from localized electrons. More interestingly, we have also observed a gate-tunable anomalous Hall effect which always emerges near the Kondo temperature. Our results show that the physical properties of LAO/STO EDLTs are dominated by the intrinsic interactions, while the LAO thickness has a secondary effect. Our experiments provide insights to understand the observed magnetism at the LAO/STO interface and demonstrate a non-trivial tuning of magnetic interactions by ionic liquid gating, paving the way for designing novel electronic devices based on LAO/STO heterostructures.

We thank Jan van Ruitenbeek, Yun-Yi Pai, Shengwei Zeng, Yu Zhang, Nikita Lebedev and Aymen Ben Hamida for useful discussions. We also thank Johannes Jobst, Hasan Atesci and Jan van Ruitenbeek for their technical support. This work is supported by the Netherlands Organisation for Scientific Research (NWO). C. Y. is supported by China Scholarship Council (CSC) with grant No. 201508110214.

\bibliographystyle{apsrev4-1}
\bibliography{LAO4}

\begin{thebibliography}{44}
\bibitem [{\citenamefont {{The molecular formula of DEME-TFSI is
  N}}()}]{textil}%
  \BibitemOpen
  \bibfield  {author} {\bibinfo {author} {\bibnamefont {{The molecular formula
  of DEME-TFSI is N}}},\ }\href@noop {} {}\bibinfo {howpublished}
  {{N-diethyl-N-methyl-N-(2-methoxyethyl)-ammonium
  bis-(trifluoromethylsulfonyl)-imide}}\BibitemShut {NoStop}%
\end{thebibliography}

\end{document}